\newcounter{myctr}

\documentclass{ws-acs}
\usepackage{url}
\usepackage{booktabs}

\begin{document}

\makeatletter
\def\@biblabel#1{[#1]}
\makeatother

\markboth{A. Antonioni and M. Tomassini}{Cooperation on Social Networks and Its Robustness}

%
\catchline{}{}{}{}{}
%

\title{COOPERATION ON SOCIAL NETWORKS\\AND ITS ROBUSTNESS}

\author{\footnotesize ALBERTO ANTONIONI\footnote{alberto.antonioni@unil.ch}\; and MARCO TOMASSINI\footnote{marco.tomassini@unil.ch}}

\address{Information Systems Department, \\ Faculty of Business and Economics, \\ University of Lausanne, \\
Lausanne, Switzerland}

\maketitle

\begin{history}
\received{(received date)}
\revised{(revised date)}
\end{history}

\begin{abstract}
In this work we have used computer models of social-like networks to show by extensive numerical simulations that cooperation
in evolutionary games can emerge and be stable on this class of networks. The amounts of cooperation
reached are at least as much as in scale-free networks but here the population model is more
realistic. Cooperation is robust with respect to different strategy update rules, population dynamics, and
payoff computation. Only when straight average payoff is used or there is high strategy or network noise
does cooperation decrease in all games and disappear in the Prisoner's Dilemma.
\end{abstract}


\section{Introduction}
\label{intro}

Game theory is used in many contexts, particularly  in economy, biology, and social sciences to describe
situations in which the choices of the interacting agents are interdependent.
 Evolutionary game theory in particular is well suited to the study of strategic interactions in animal and human populations and
has a well defined mathematical structure that allows analytical conclusions to be reached in the
context of well-mixed and very large populations 
(see e.g.~\cite{weibull95,Hofbauer1998}). However, starting with the works of Axelrod~\cite{axe84}, and Nowak and May~\cite{nowakmay92}
population structures with local interactions have been brought to the focus
of research since they are supposed to be closer model of real social structures. Indeed, social interactions can be more precisely
represented as networks of contacts in which nodes represent agents and links stand for their 
relationships. Thus, in the wake of the flurry of research on complex networks~\cite{newman-book}, evolutionary game theory has been recently enhanced with networked population structures.
The corresponding literature has already grown to a point that makes it difficult to be exhaustive; however,
good recent reviews can be found in~\cite{szabo,anxo1,mini-rev}.
In particular, the problems of cooperation and coordination, leading to social efficiency problems and social dilemmas, have been
studied in great detail through some well known paradigmatic games such as the Prisoner's Dilemma, Snowdrift, and
the Stag Hunt. Those games, although unrealistic with respect to actual complex human behavior, constitute a simple way
for gaining a basic understanding of these important issues. One of the important findings in this line of research is that the heterogeneous structure
of many model and actual networks may favor the evolution of socially valuable equilibria, for instance leading to non-negligible amounts
of cooperation in a game such as the Prisoner's Dilemma where the theoretical result should be generalized 
defection~\cite{santos-pach-05,santos-pach-06,anxo1}.
Coordination on the socially preferable Pareto-efficient equilibrium in the Stag Hunt and an increase of the fraction of cooperators
in the Snowdrift game have also been observed on populations interacting according to a complex network structure~\cite{santos-pach-06,anxo1}.

Most detailed results on evolutionary games on networks to date have been obtained for network types that are standard in 
graph theory such as Erd\"os-R\'enyi random graphs, scale-free
networks, Watts--Strogatz small-world networks, and a few others that are commonly used (see~\cite{anxo1} for a good complete review). 
This is obviously an important first step, since general results have been obtained for these network topologies, especially through numerical
simulations and in some cases even analytical ones~\cite{szabo}.
However, since the theory is of interest
especially when applied in human or animal societies, to move a further step on, the network structures used should be as close as possible to those that can be observed in such contexts. This is the realm of social networks, many of which have been investigated and their main
statistics established (see, for instance,~\cite{newman-book}). While social networks degree distributions have been found to
be broad-scale in general and thus they share this property~\cite{am-scala-etc-2000}, at least in part, with model scale-free networks they also
possess some features that are not found in most common model networks. The three more important for our purposes
are: clustering, positive degree assortativity, and community structure~\cite{newman-03,newman-book}. Thus, in the present study
we would like to offer a rather systematic study of evolutionary games played on this kind of networks in order to pave the way
for the understanding of their behavior in real societies.
Actually, there have already been several studies of evolutionary games on particular social networks in the 
past (see, for example and among others,~\cite{trusina-karate-club,lozano-plos,luthi-pest-tom-physa08}). Broadly speaking, 
these investigations all tend to show that socially valuable outcomes such as cooperation
and coordination are more likely to evolve than in unstructured populations. However, the particular nature of the
few networks used, although it provides some insight, does not allow one to draw more general and statistically valid
conclusions. For this reason, here, instead of studying another particular network, we prefer to follow the methodology
used in investigations such as~\cite{santos-pach-06,anxo1}, where numerical simulation results are validated by using
many runs on different networks belonging to the same class. As a model for constructing social networks, among several different possibilities,
we choose to use Toivonen et al.'s model~\cite{toivonen-2006}, which will be discussed in Sect.~\ref{networks}.
This model has already been investigated in~\cite{luthi-pest-tom-physa08}, albeit for a different evolution rule and for a reduced game phase space. 
In this paper, in addition to numerically studying the steady states
of evolutionary games on these social network models when using static networks and error-free strategy update rules,
we shall also briefly explore the effect of errors on strategy update rules and noise on network structure.

The paper is organized as follows. We first present some background on two-person, two-strategies symmetric games, followed by
evolutionary games on networks. This is followed
by a description of the model networks used. Next, results for the chosen games are presented and discussed, as well as the effects induced
by the introduction of noise. We then offer our conclusions and some ideas for future work.

\section{Games Studied}
\label{games}

We have studied the four classical two-person, two-strategies symmetric games described by the generic payoff bi-matrix of Table~\ref{pbm}. 
\begin{table}[!ht]
\begin{center}
{\normalsize
$
\begin{array}{c|cc}
 & C & D\\
\hline
C & (R,R) & (S,T)\\
D & (T,S) & (P,P)
\end{array}
$}
\caption{Generic payoff bi-matrix for the two-person, two-strategies symmetric games.
$C$ and $D$ are the possible strategies, and $R,T,P$, and $S$ are utility values as
discussed in the text. \label{pbm}}
\end{center}
\end{table}
In this matrix, $R$ stands for the \textit{reward}
the two players receive if they
both cooperate ($C$), $P$ is the \textit{punishment} for bilateral defection ($D$), and $T$  is the
\textit{temptation}, i.e.~the payoff that a player receives if she defects while the
other cooperates. In the latter case, the cooperator gets the \textit{sucker's payoff} $S$. Payoff values may undergo any affine transformation
without affecting neither the Nash equilibria, nor the dynamical fixed points; however, the parameters' values are restricted to the
``standard''  configuration space
defined by $R=1$, $P=0$, $-1 \leq S  \leq 1$, and $0 \leq T \leq 2$. In the resulting $TS$ plane, each game's space corresponds
to a different quadrant depending on the ordering of the payoffs.

If the payoff values are ordered such that
$T > R > P > S$ then
defection is always the best rational individual choice, so that 
$(D,D)$ is the unique Nash Equilibrium (NE) and also the only Evolutionarily Stable Strategy
(ESS)~\cite{weibull95}. This famous game is called \textit{Prisoner's Dilemma} (PD).
Mutual cooperation  would be socially preferable but $C$ is strongly dominated by $D$. 

In the \textit{Snowdrift} (SD) game, the order of $P$ and $S$ is reversed, yielding $T > R > S > P$. Thus, in the SD 
when both players defect they each get the lowest payoff.
$(C,D)$ and $(D,C)$ are NE of the game in pure strategies. There is
a third equilibrium in mixed strategies which is the only dynamically stable state,
while the two pure NE are not~\cite{weibull95}.
Players have a strong incentive
to play $D$, which is harmful for both parties if the outcome produced happens to be $(D,D)$.

With the ordering  $R > T > P > S$ we get the \textit{Stag Hunt} (SH) game in which mutual cooperation $(C,C)$ is the best outcome,
Pareto-superior, and a NE.  The second NE, where both players defect
is less efficient but also less risky. The dilemma is represented by the fact that the
socially preferable coordinated equilibrium $(C,C)$ might be missed for ``fear'' that the other player
will play $D$ instead. 
The third mixed-strategy NE in the game is evolutionarily unstable~\cite{weibull95}. 

Finally, the \textit{Harmony}  game has $R > S > T > P$ or $R>T>S>P$. $C$ strongly dominates $D$ and
the trivial unique NE is $(C,C)$. This game is non-conflictual by definition and does not cause any
dilemma: it is included just to complete the quadrants of the parameter space.

With the above conventions, in the figures that follow, the PD space is the lower right quadrant; the SH is the
lower left quadrant, and the SD is in the upper right one. Harmony is represented by the upper left
quadrant.

\section{Evolutionary Games on Networks}
In this section we present background material on evolutionary games on finite-size populations of agents represented
by networks of contacts to make the paper as self-contained as possible.

 \subsection{Population structure} 
The population of players is represented by a connected unweighted, undirected  graph $G(V,E)$, where the
set of vertices $V$ represents the agents, while the set of edges  $E$ represents their symmetric interactions. The
 population size $N$ is $|V]$, the cardinality of $V$. The set of neighbors of an agent $i$ is defined as: $V_i =\{j \in V\: |\: \mathit{dist}(i,j)=1\}$, 
and its cardinality $|V_i|$ is the degree $k_i$ of vertex $i \in V$. The average
degree of the network is called $\bar k $ and $P(k)$ denotes its degree distribution function, i.e. the probability
that an arbitrarily chosen node has degree $k$. The network topologies used are explained in Sect.~\ref{networks}.

\subsection{Payoff calculation and strategy revision rules}
\label{payoff}

In evolutionary game theory, one must specify how individual's payoffs are computed and how
agents revise their present strategy. In the standard theory, there is a very large well-mixed population;
however, when the model is applied to a finite population whose members are the vertices of a graph, each agent $i$
can only interact with agents contained in the neighborhood $V_i$, i.e. only local interactions are permitted.

\noindent Let $s_i \in \{C,D\}$ be the current strategy of player $i$ and 
let us call $M$ the payoff matrix of the game. The quantity
$$\Pi_i(t) =  \sum _{j \in V_i} \sigma_i(t)\; M\; \sigma_{j}^T(t)$$
\noindent is the \textit{accumulated payoff} collected by agent $i$ at time step $t$ and $\sigma_i(t)$ is  a vector
giving the strategy profile at time $t$ with $C= (1 \;\; 0)$ and $D = (0 \;\; 1)$. We also use the \textit{average payoff} $\overline{\Pi}_i (t)$ defined as the average of
accumulated payoff collected by a given agent $i$ at time step $t$:
$$\overline{\Pi}_i (t) = \frac{1}{k_i}\: \sum _{j \in V_i} \sigma_i(t)\; M\; \sigma_{j}^T(t)$$

Several strategy update rules are commonly used.
Here we shall describe three rules belonging to the \textit{imitative} class that have been used in our simulations; the first rule
is deterministic, while the following strategy updates are stochastic.

\noindent The first rule 
is to switch to the strategy of the neighbor that has scored
best in the last time step. This \textit{imitation of the best} policy can be described in the following way:
the strategy $s_i(t)$ of individual $i$ at time step $t$ will be
$$s_i(t) = s_j(t-1),$$
where
$$j \in \{V_i \cup i\} \;s.t.\; \Pi_j = \max \{\Pi_k(t-1)\}, \; \forall k \in \{V_i \cup i\}.$$
\noindent That is, individual $i$ will adopt the strategy of the player with the highest
payoff among its neighbors including itself.
If there is a tie, the winner individual is chosen uniformly at random, but otherwise the rule is deterministic.

\noindent The \textit{local replicator dynamics} rule is stochastic~\cite{hauer-doeb-2004}.  Player $i$'s strategy $s_i$ is updated by drawing
another player $j$  from the neighborhood $V_i$ with uniform probability,
and replacing $s_i$ by $s_j$ with probability: 
$$p(s_i \rightarrow s_j) = (\Pi_j - \Pi_i)/K,$$
if $ \Pi_j > \Pi_i$, and keeping the same strategy if  $ \Pi_j \le \Pi_i$. $K=\max(k_i,k_j)[(\max(1,T)-\min(0,S)]$, with $k_i$ and $k_j$ being the
degrees of nodes $i$ and $j$ respectively, ensures
proper normalization of the probability $p(s_i \rightarrow s_j)$.

\noindent The last strategy revision rule is the \textit{Fermi rule}~\cite{szabo}: 
$$p(s_i \rightarrow s_j) =\frac{1} { 1+ \exp(-\beta(\Pi_j - \Pi_i))}.$$
This gives the probability that player $i$ switches from strategy $s_i$ to $s_j$, where $j$ is a randomly
chosen neighbor of $i$. $\Pi_j -\Pi_i$ is the difference of payoffs earned by $j$ and $i$ respectively.
The parameter $\beta$ in the function gives the amount of noise: a low $\beta$ corresponds to high probability of error and, conversely,
high $\beta$ means low error rates. This interpretation comes from physics, where the reciprocal of $\beta$ is called
the temperature. Consequently, payoffs will be more noisy as temperature is raised ($\beta$ is lowered).
In the above expressions we have used the accumulated payoff $\Pi_i$. Analogous formulae hold for average payoff $\overline{\Pi}_i$.

\subsection{Strategy update timing}
\label{timing}

Usually, agents systems in evolutionary game theory  are updated synchronously. However, strictly
speaking, simultaneous update is
physically unfeasible as it would require a global clock,
while real extended systems in biology and society in general have to take into account finite signal propagation speed.
Furthermore, simultaneity may cause some artificial effects in the dynamics which are not observed
in real systems~\cite{hubglance93}. However, for evolutionary game theory on networks the common wisdom is that
the timing of updates does not influence the system properties in a fundamental manner and results
are similar in most cases~\cite{anxo1,Grilo2011109}, with asynchronism being sometimes beneficial to the
emergence of cooperation~\cite{Grilo2011109}. This is called an elementary time step.
To compare synchronous and asynchronous dynamics, here we use the customary
fully asynchronous update, i.e.~updating a randomly chosen agent at a time with 
replacement. The two dynamics, synchronous and fully asynchronous are the extremes cases.
It is also possible to update the agents in a partially synchronous manner where
a fraction $f =n/N$ of randomly chosen agents is updated in each time step.
 When $n = N$ we recover the fully synchronous update, while
$n=1$ gives the extreme case of the fully asynchronous update. Varying $f$ thus allows one to investigate the role
of the update policy on the dynamics~\cite{luthi-pest-tom-dyn,Grilo2011109}.

 \subsection{Simulation parameters}       
 
The networks used in all simulations are of size $N=4000$ with mean degree $\bar k =8 $.
 The $TS$ plane has been sampled with a grid step of $0.05$ and
 each value in the phase space reported in the figures is the average of $100$ independent runs, using a fresh graph 
 realization for each run. The initial graph for each run doesn't change in the static case, while it evolves
 in the dynamic case, as described in the main text.
 Note that steady states have always been reached when strategies evolve on a static graph.  
 In the asynchronous dynamics, we first let the system evolve for a transient period of $4000 \times N \simeq 16 \times 10^6$ elementary time steps.
 In the synchronous case, the same total number of updates is performed.
 The averages are calculated at the steady state that is reached after the transient period. 
 True equilibrium states in the sense of stochastic stability are not guaranteed
 to be reached by the simulated dynamics. For this reason we prefer to use the terms steady states
 which are configurations that have little or no fluctuation over an extended period of time.  
 In the case of fluctuating networks, the system as a whole never reaches a steady state in the sense specified above.
 This is due to the fact that the link dynamics remains always active. However, the distribution of
 strategies on the network does converge to a state that shows little fluctuation, i.e. a steady state.

\section{Network Construction and Properties}
\label{networks}

\begin{figure} [!htb]
\begin{center}
\resizebox{1\textwidth}{!}{%
\begin{tabular}{c}
 \mbox{\includegraphics[width=0.8cm]{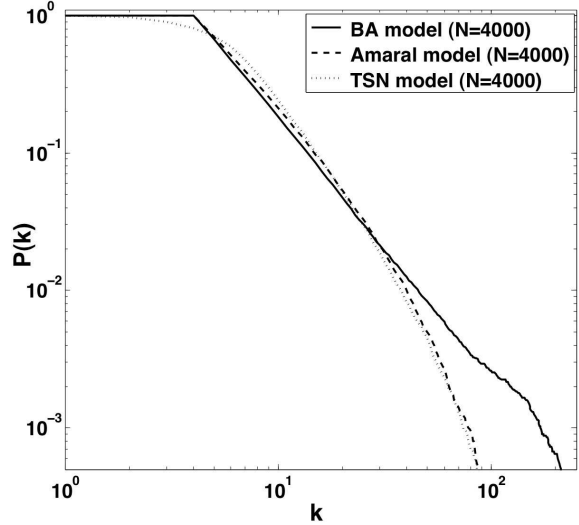} } \protect 
\end{tabular}}
\end{center}
\vspace{-1.2cm}
\caption{Cumulated degree distribution of a TSN  compared with a standard Barab\'asi--Albert network
of the same size and mean degree (see text).\label{degree}}
\end{figure}

There exist several models for constructing social-like networks~\cite{Toivonen2009240}. Among them, we have chosen
Toivonen et al.'s model~\cite{toivonen-2006}, called here the Toivonen Social Network (TSN), which 
 was conceived to construct a network with most
of the desired features of real-life social networks i.e., assortativity, high clustering coefficient, community structure, having an adjustable decay rate of the degree distribution, and a finite cut-off.
The TSN construction and properties are described in detail in~\cite{toivonen-2006}. The process we have used to obtain a TSN can be summarized as follows:
\begin{enumerate}
\item Start with a small clique formed by $m_0$ vertices.
\item A new vertex $v$ is added to the network and it is connected to $m_r$ vertices chosen with uniform probability in the existing network.  (\emph{random attachment})
\item Vertex $v$ is connected to $m_s$ vertices chosen with uniform probability within the list of neighbors of its neighbors. (\emph{implicit preferential attachment}) \\ 
Every time that vertex $v$ is connected to a new vertex the list of neighbors of its neighbors is updated. 
\item Repeat step 2 and 3 until the network has grown to desired size $N$.
\end{enumerate}
Notice that the process responsible for the appearance of high clustering and community structures is step 3. Moreover, this model is slightly different
from the original one, because in the latter the list of neighbors in step 3 is not updated as in the former. This modification does not change
the original model proprieties. \\
In the simulations, we have used networks of size $N=4000$ with a clique of $m_0=9$ initial nodes as starting network.
Every time a new node is added at step 2, its number of initial contacts $m_r$ is distributed with probability: $\mathbb{P}(m_r=1)=0.5$ and $\mathbb{P}(m_r=2)=0.5$, while 
the number of its secondary contacts $m_s$ is uniformly distributed between, and including, the integer values $0$ and $5$. The resulting degree distribution falls below a power law as shown 
in~\cite{toivonen-2006}. The average degree of the network is given by the formula that provides the average degree of a new added vertex:
$$1\cdot\mathbb{P}(m_r=1) + 2\cdot\mathbb{P}(m_r=2) + \sum_{t=0}^5 t \cdot\mathbb{P}(m_s=t) = 4$$
Thus, the resulting average degree of the network is $\bar{k} = 2\cdot 4 = 8$.

To compare with another complex network model besides the standard Barab\'asi--Albert (BA) networks~\cite{barabasi-sf}, we have
also used a model proposed by Amaral et al.~\cite{am-scala-etc-2000} which also features several of the characteristics
of real complex networks, especially the distribution tail cutoff, and takes into account aging and cost constraints in the pure preferential attachment BA model.

In Fig.~\ref{degree} we depict the degree distribution of a realization of TSN compared with a Barab\'asi--Albert scale-free 
network, and the Amaral et al. model
(in all cases, $N=4000$ and $\bar k=8$). 
The degree distribution of both TSN and the Amaral et al. model falls down faster than in a BA scale-free network and they are
 in fact closer to a stretched exponential or an exponentially truncated power-law. This
agrees with many observations on empirically measured social networks, see e.g.~\cite{am-scala-etc-2000,newman-collab-1,tom-leslie-evol-net-07}.

Table~\ref{tab-newman} collects a number of statistics on social networks, mainly collaboration networks in various 
scientific disciplines. These empirical data
show that, irrespective of the network size, they are all degree-assortative ($r>0$) and have a high clustering coefficient. These features are
shared by TSNs, and partly by the Amaral model, but not by BA networks. Indeed, it has been shown in the literature that most social networks possess these 
features (for a review, see~\cite{newman-03}).

\begin{table}[!htb]
\small
\begin{center}
\vspace{3mm}
\begin{tabular}{lrrrrc} \toprule
\multicolumn{1}{}{}&\multicolumn{1}{c}{N}&\multicolumn{1}{c}{$\bar k$}&\multicolumn{1}{c}{C}&\multicolumn{1}{c}{r}&\multicolumn{1}{c}{Ref.} \\ \cmidrule(r){2-6}
Biology   					& 1 520 251 &  15.53 & 0.60 & 0.127 & \cite{newman-03} \\
Mathematics   					& 253 339 &  3.92 & 0.34 & 0.120 & \cite{newman-03}\\
Physics   					& 52 909 &  9.27 & 0.56  & 0.363 &  \cite{newman-03} \\
Computer Science		& 11 994 & 2.55  & 0.50 & - & \cite{newman-collab-1} \\
Genetic Programming (GP)		& 2 809 &  4.17  & 0.66 & 0.130 & \cite{tom-leslie-evol-net-07} \\
TSN model			& 4 000 &  $\sim$ 8.00 &  0.36 & 0.123 & - \\
Amaral model			& 4 000 & 8.00 & 0.22 & -0.193 & - \\
BA scale-free networks 	& 4 000 &  8.00 & 0.01 & -0.042 & - \\
\bottomrule
\end{tabular}
\caption{Some statistical features of model and real networks of coauthorships. $N$ is the network size (number of nodes). Average degree $\bar k$; Average clustering
coefficient $C=\frac{1}{N} \sum_{j=1,N} C_i$, where $C_i$ is the node's individual clustering coefficient~\cite{newman-03}; $r$ is the Pearson's coefficient of
neighbors' degree correlation~\cite{newman-03}.
}
\label{tab-newman}
\end{center}
\end{table}

\begin{figure} [!htb]
\begin{center}
\resizebox{1\textwidth}{!}{%
\begin{tabular}{c}
\mbox{\includegraphics[width=1.2cm] {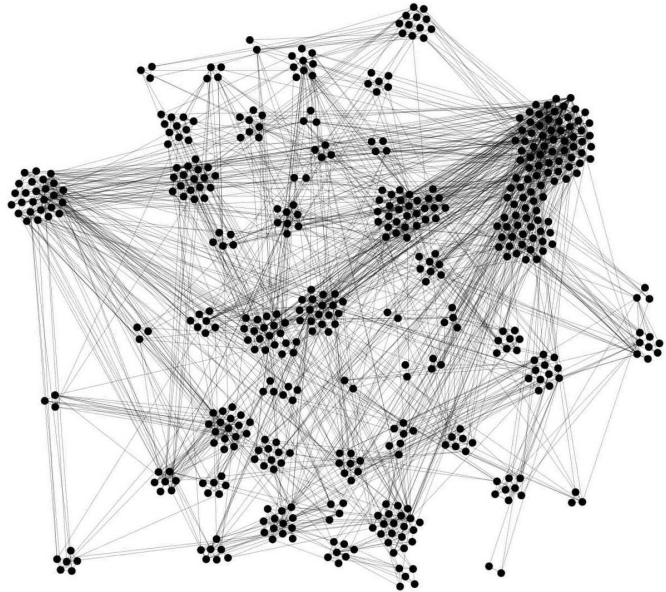} } \protect
\end{tabular}}
\end{center}
\vspace{-1.2cm}
\caption{Community structure in a small TSN ($N=500$). \label{toivonen}}
\end{figure}

Finally, another characteristic feature of most social networks is the presence of communities, i.e. roughly speaking, groups of nodes
that are more interconnected in the group than between groups~\cite{newman-book}. TSNs do reproduce this feature while models
such as the BA do not. For the purpose of illustration only, Fig.~\ref{toivonen} shows a small example ($N=500$) of a TSN graph in which communities
have been detected using an algorithm based multi-level optimization of modularity~\cite{blondel}. Indeed, its modularity 
score~\cite{newman-06} is $0.544$ confirms the presence of communities, although to assess its statistical significance
would require proper comparison with a suitable null model. Amaral model also gives rise to strongly connected clusters
of nodes and its modularity is correspondingly very high, about $0.87$.

In conclusion, we shall mainly use the TSN model for generating a family of networks, hoping to be able to reproduce the main features
of actual social networks. The Amaral model will also be used in part to compare results of evolutionary games in a couple of
more realistic models besides the standard BA one.

\section{Results}

In this section we report and discuss our simulation results for the $TS$-plane using different
strategy update rules, network topology, and evolution dynamics.

\subsection{Replicator dynamics}

\begin{figure} [!htb]
\begin{center}
\resizebox{1\textwidth}{!}{%
\begin{tabular}{c}
 \mbox{\includegraphics{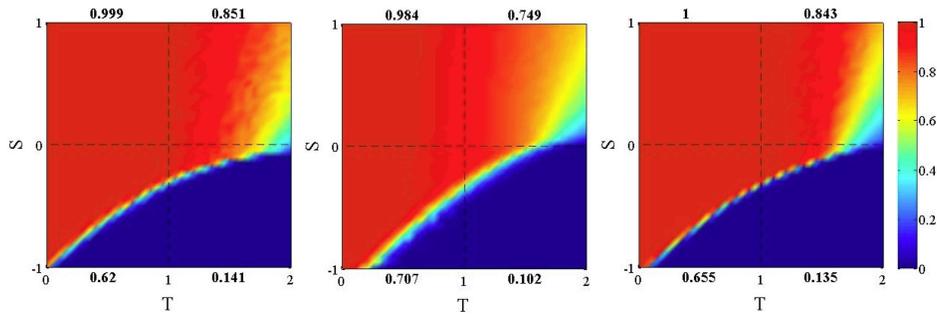} } \protect 
\end{tabular}}
\end{center}
\caption{Asymptotic distribution of strategies in the TS plane in BA networks (left), Amaral's model (middle), and 
TSNs (right) using replicator dynamics and accumulated payoff.
 Initial density of cooperators is $0.5$ uniformly distributed at random in all cases. Population evolution
 is asynchronous. Network size is $N=4000$ and
 average degree $\bar k=8$ in both cases. Values are averages over $100$ independent runs. The numbers
 in bold next to each quadrant stand for the average cooperation in the corresponding game.\label{rd-async}}
\end{figure}

Figure~\ref{rd-async} shows average cooperation frequencies for the four games in the case of
BA networks (left image), Amaral's networks (middle), and TSNs (right image). The payoff is accumulated payoff (see Sect.~\ref{payoff}).
For the BA case, the results are very similar to those
obtained in~\cite{santos-pach-06,anxo1} for larger systems with the same mean degree.
Both visually and from the average 
game cooperation values shown next to each quadrant, it is clear that TSNs  and the Amaral model networks are as favorable to
cooperation as the idealized BA case when the strategy update rule is local replicator dynamics.
However, the mechanisms responsible for high cooperation seem to be different. While in BA networks highly connected cooperator 
hubs play the role of catalyzers for the diffusion of the strategy~\cite{santos-pach-05}, in social-like networks cooperation may thrive thanks
to the higher clustering coefficient and the presence of communities. When a local cluster becomes colonized by cooperators, it tends to
be robust against attacks by defectors~\cite{anxo1}.

The results shown are for asynchronous update (see Sect.~\ref{timing}), results for the synchronous
case are very similar and we do not show them. The results on TSNs also confirm those obtained 
in~\cite{luthi-pest-tom-physa08} where a different  update rule and parameter space was used, while the
BA case was first shown in~\cite{santos-pach-06}.

\begin{figure} [!htb]
\begin{center}
\resizebox{1\textwidth}{!}{%
\begin{tabular}{c}
 \mbox{\includegraphics{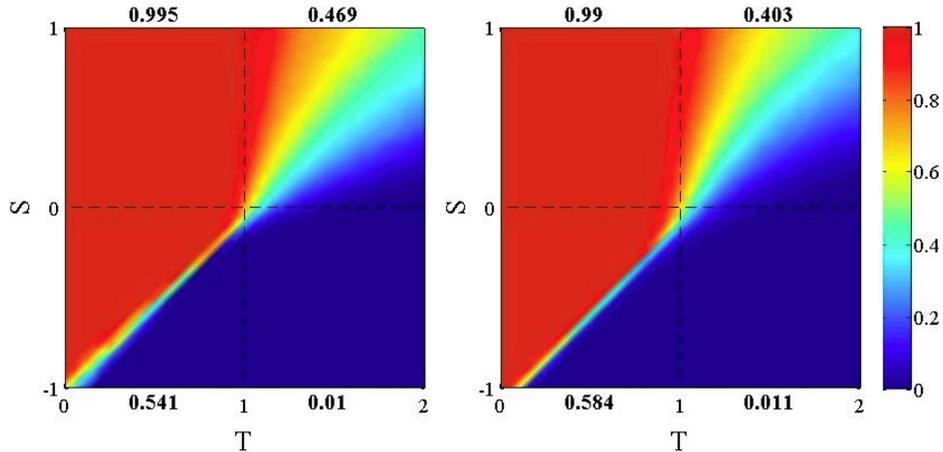} } \protect 
\end{tabular}}
\end{center}
\caption{Asymptotic distribution of strategies in the TS plane in BA networks (left) and TSNs (right) using replicator dynamics
and average payoff.
 Initial density of cooperators is $0.5$ uniformly distributed at random in all cases. Network size is $N=4000$ and
 average degree $\bar k=8$ in both cases. Asynchronous population dynamics. Values are averages over $100$ independent runs.  \label{rd-av-asy}}
\end{figure}

In Fig.~\ref{rd-av-asy} we plot average cooperation levels in all games for the case in which agent's payoffs are
averaged over the agent's links (see Sect.~\ref{payoff}). As expected, using average payoffs instead of accumulated ones
corresponds to make the network more degree-homogeneous and thus the results resemble those obtained
in regular random graphs~\cite{anxo1} and are similar for both BA networks and TSNs. Defection becomes almost
complete in the PD, values are similar to the well-mixed case in the SD, and in the SH the usual bistable
result is recovered.

We conclude with~\cite{santos-average,Masuda2007,Tomassini2007a,attila-average}
that the use of a mean-field for the payoff scheme is detrimental to cooperation. Whether or not this significant factor
in social networks is difficult to assess but, obviously, beyond a certain limit, maintaining links becomes expensive
in practice and furthermore their frequency of usage should decrease. Probably, real situations are somewhere in between these
two extreme cases and thus it is interesting to see what happens when average and accumulated payoff computations
both happen to some extent. 
\begin{figure} [!htb]
\begin{center}
\resizebox{1\textwidth}{!}{%
\begin{tabular}{c}
 \mbox{\includegraphics[width=0.8cm]{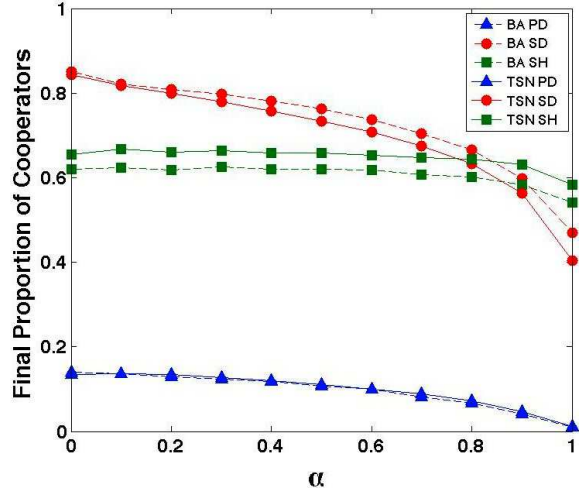} } \protect 
\end{tabular}}
\end{center}
\vspace{-1.2cm}
\caption{Final average proportion of cooperation in the PD, SD, and SH games as a function of $\alpha$ (see text) for
BA and TSN networks. The update is asynchronous and the revision rule is replicator dynamics.  \label{acc-aver-rd}}
\end{figure} 
Finally, following the idea of Szolnoki et al.~\cite{attila-average}, we compute payoffs using a weight parameter $\alpha \in [0,1]$ to
represent the proportion of average payoff. Thus,
the actual payoff to agent $i$ is given by the formula: $\alpha \overline{\Pi}_i + (1-\alpha) \Pi_i$.
In Fig.~\ref{acc-aver-rd} we plot the steady-state average cooperation frequency for several $\alpha$ values. Clearly, in all games
cooperation decreases with increasing $\alpha$ but the losses are limited in absolute value, except for the SD. Nevertheless, compared
with average cooperation at $\alpha=0$, which is $0.135$ for TSN and $0.141$ for BA,  the PD looses all residual cooperation.

\subsection{Imitation of the best}

\begin{figure} [!htb]
\begin{center}
\resizebox{1\textwidth}{!}{%
\begin{tabular}{c}
 \mbox{\includegraphics{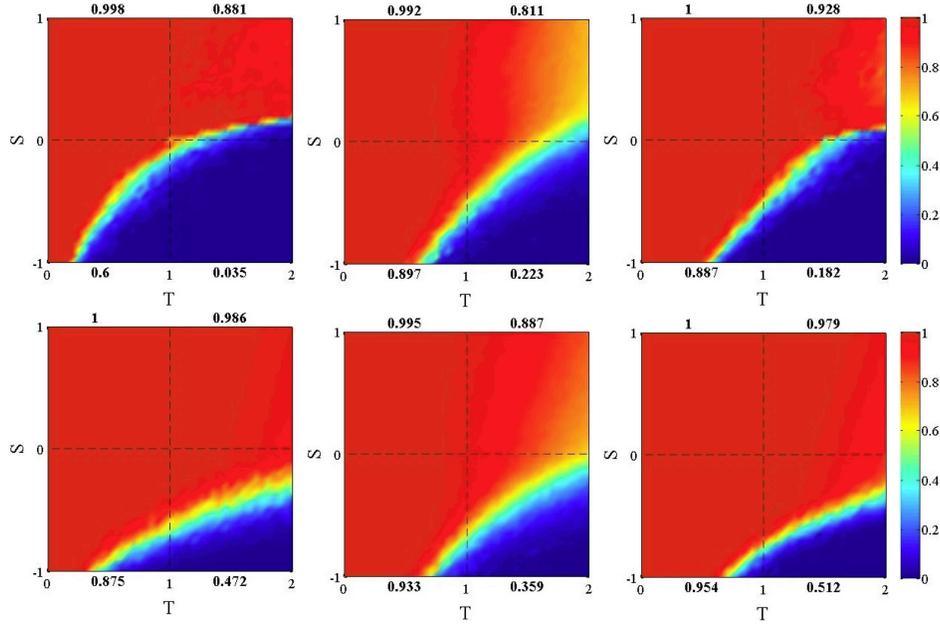} } \protect 
\end{tabular}}
\end{center}
\caption{Asymptotic distribution of strategies in the TS plane in BA networks (left column), Amaral's networks (middle), 
and TSNs (right column) using 
imitation of the best and accumulated payoff. Top row: synchronous dynamics. Bottom row: asynchronous dynamics.
 Initial density of cooperators is $0.5$ uniformly distributed at random in all cases. Network size is $N=4000$ and
 average degree $\bar k=8$ in both cases.
 Values are averages over $100$ independent runs. \label{ib}}
\end{figure}

In this section we compare BA networks with TSNs, and the Amaral model using synchronous and asynchronous population dynamics and
the imitation of the best strategy update rule (see Sect.~\ref{payoff}). The average cooperation results with accumulated payoff
are illustrated in Fig.~\ref{ib}.
Note that asynchronous dynamics (bottom figures) is more favorable for cooperation in this case for all topologies, as already remarked
by Grilo and Correia~\cite{Grilo2011109} for the BA case. With this deterministic asynchronous dynamics TNSs are
particularly conducive to cooperation, more than $0.5$ in the average for the PD, $0.98$ in the SD phase space, and $0.95$
in the SH quadrant; cooperation is a bit lower for the PD  for the Amaral's case but still good in the whole game phase
space. The qualitative reason for the diffusion and stability of
cooperation seems to be related to the higher clustering coefficients of TSNs with respect to BA networks (see Table~\ref{tab-newman}).
In these networks, if a tightly linked cluster happens to get a large majority of cooperators, it can spread cooperation more easily to
next neighbors. This is also true in the 
synchronous case (top figures) but to a lesser extent for the PD, and almost to the same extent for the SD and the SH, .

\begin{figure} [!htb]
\begin{center}
\resizebox{1\textwidth}{!}{%
\begin{tabular}{c}
 \mbox{\includegraphics{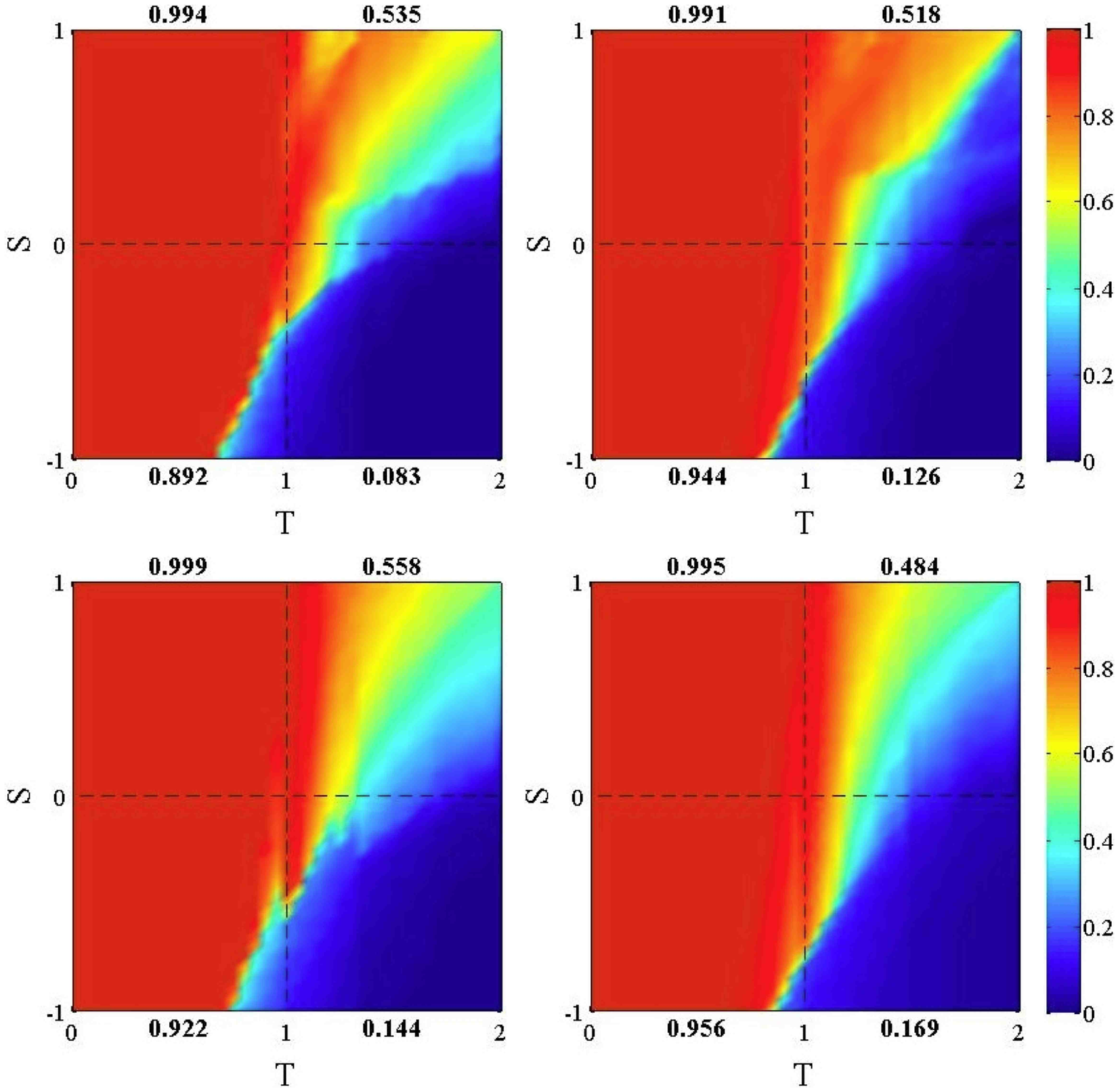} } \protect 
\end{tabular}}
\end{center}
\caption{Asymptotic distribution of strategies in the TS plane in BA networks (left column) and TSNs (right column) using 
imitation of the best and average payoff. Top row: synchronous dynamics. Bottom row: asynchronous dynamics.
 Initial density of cooperators is $0.5$ uniformly distributed at random in all cases. Network size is $N=4000$ and
 average degree $\bar k=8$ in both cases. 
 Values are averages over $100$ independent runs. \label{ib-average}}
\end{figure}

The results with average payoff for the BA model and TSNs again with synchronous and asynchronous population
dynamics are shown in Fig.~\ref{ib-average}. 
The top row shows images for synchronous update in BA networks
(left image) and TSNs (right image). When using fully averaged payoffs, the results become similar to those
obtained in a regular random graph with the same average degree, as in~\cite{anxo1}, with a small
advantage in cooperation for the TSNs. In the asynchronous case (bottom row) results are similar with somewhat
more cooperation in both cases and again an advantage for the TSNs. The results for BA networks have already been
obtained by Grilo and Correia and are consistent with ours~\cite{Grilo2011109}.

\begin{figure} [!htb]
\begin{center}
\resizebox{1\textwidth}{!}{%
\begin{tabular}{c}
 \mbox{\includegraphics[width=0.8cm]{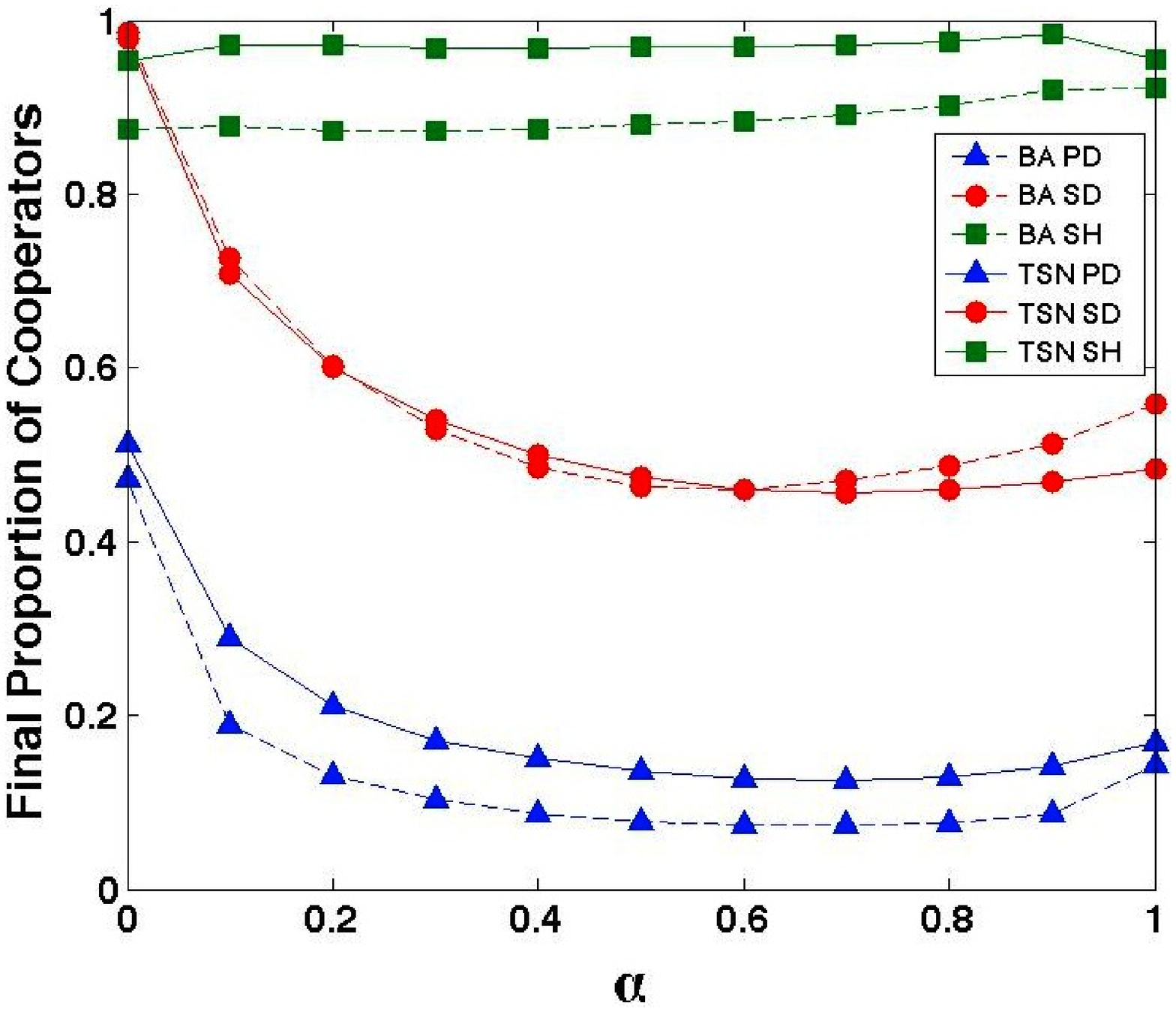} } \protect 
\end{tabular}}
\end{center}
\vspace{-1.2cm}
\caption{Final average proportion of cooperation in the PD, SD, and SH games as a function of $\alpha$ (see text) for
BA and TSN networks. Population update is asynchronous. The strategy revision rule is imitation of the best. \label{medie-ib}}
\end{figure}

Finally, we present in Fig.~\ref{medie-ib} the average amount of cooperation for the asynchronous
dynamics as a function of the parameter $\alpha$ as explained in the previous section.
The most important remark is that cooperation in the TSN model is almost always higher in the whole range of $\alpha$ for all games
with respect to BA networks. 
Concerning the average/accumulated payoff tradeoff, one can see that there is a large decrease of cooperation
in the SD and in the PD games going from pure accumulated ($\alpha=0$) to pure average ($\alpha=1$). The SH is
less affected. 

\subsection{Fermi rule}

\begin{figure} [!htb]
\begin{center}
\resizebox{1\textwidth}{!}{%
\begin{tabular}{c}
 \mbox{\includegraphics{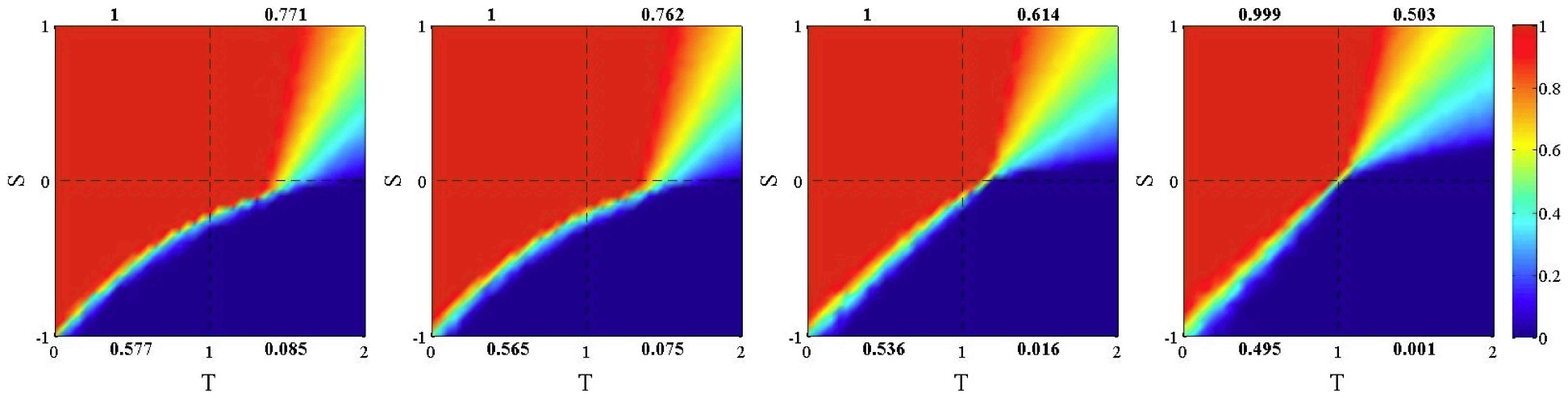} } \protect 
\end{tabular}}
\end{center}
\caption{Asymptotic distribution of strategies in the TS plane in TSNs  using the Fermi rule with
accumulated payoff and synchronous dynamics.
 Initial density of cooperators is $0.5$ uniformly distributed at random in all cases. Network size is $N=4000$ and
 average degree $\bar k=8$. From left to right: $\beta=10,1, 0.1, 0.01$. Values are averages over $100$ independent runs. \label{fermi}}
\end{figure}

The Fermi strategy update rule allows for the introduction of random noise in the choice of strategy (see Sect.~\ref{payoff}).
In Fig.~\ref{fermi} we show average cooperation results for different $\beta$ values, from left to right $\beta=10,1, 0.1, 0.01$.
The population dynamics is synchronous and thus the results can be compared with the analogous ones obtained by 
Roca et al.~\cite{anxo1} in the case of BA networks. The results are quite similar, with slightly less cooperation in TSNs, and one can 
see that average cooperation levels
tend to decrease with decreasing $\beta$ since this corresponds to increasing randomness in the choice of strategy.

The asynchronous case is very similar and is not shown to save space. Clearly, those heterogeneous networks make
cooperation relatively robust when there is
some noise in the choice of strategies, but when the decision becomes almost random topological considerations do not
play an important role any longer.

\subsection{Network links fluctuations}

In a recent study, we have shown by numerical simulations that a certain amount of link rearrangement during
evolution may hamper cooperation from emerging in scale-free networks~\cite{antonioni}. It is important to note
that the dynamics we are referring to is not strategy-related as in other work 
(see, for instance,~\cite{zimm-et-al-04,santos-plos-06,luthi-pest-tom-dyn,mini-rev}); 
rather, noise is exogenous
and random. Under these conditions, we found in~\cite{antonioni} that the maintenance of neighborhoods and clusters becomes
increasingly difficult with growing noise levels, and this has a negative effect on cooperation. In this section, we shall investigate the same issue for TSNs. 

In actual social networks, both the number of links and of nodes tend to increase in time with a speed that
is characteristic of a given network, and link formation is not random (see e.g.~\cite{kossi-watts-06,tom-leslie-evol-net-07} for
two empirical analyses of growing social networks). However,  
little is known about these complex dynamics  and we are not aware of any general theoretical model.

To test for the robustness of cooperation under very simple hypotheses on network fluctuations, and following~\cite{antonioni}, 
we have designed a link dynamics well adapted to TSNs through link cutting and rewiring that works as follows:

\begin{itemize}
\item a vertex $v$ is chosen uniformly at random in $V$ and its degree $k$ is stored
\item \textbf{cutting edges to neighbors}: node $v$ looses all links to its neighbors, except to those that would become isolated by doing so
and whose number is $k_1$
\item \textbf{relinking node v}: vertex $v$ creates $(k-k_1)$ new links:
\begin{enumerate}
\item node $v$ connects itself uniformly at random  with another node $u \notin V_v$
\item vertex $v$ is connected to $4$ vertices chosen with uniform probability within the list of neighbors of its neighbors. 
Every time that vertex $v$ is connected to a new vertex the list of neighbors of its neighbors is updated 
\item Repeat step 1 and 2 until the number of new links equals $(k-k_1)$
\end{enumerate}
\end{itemize}

This kind of rewiring could be interpreted as a sort of \textit{migration} of node $v$. In the cutting step all of $v$'s  previous links are
suppressed and the same number of links is created in the relinking step in another region of the network. Vertex $v$ keeps its
original degree and thus the mean degree stays constant. If the ``moving'' node $v$ had neighbors of degree
one, those will follow it in the new position to avoid isolated nodes.  
The degree distribution function changes only slightly, the clustering coefficient remains high, above $0.30$, and
the network is always degree-assortative, all of
which is consistent with empirical observations on real social networks~\cite{kossi-watts-06,tom-leslie-evol-net-07}. Note, however, 
that our dynamics represents only
one among many reasonable possibilities. Reasons of space prevent us from studying other models here.

\begin{figure} [!htb]
\begin{center}
\resizebox{1\textwidth}{!}{%
\begin{tabular}{c}
 \mbox{\includegraphics{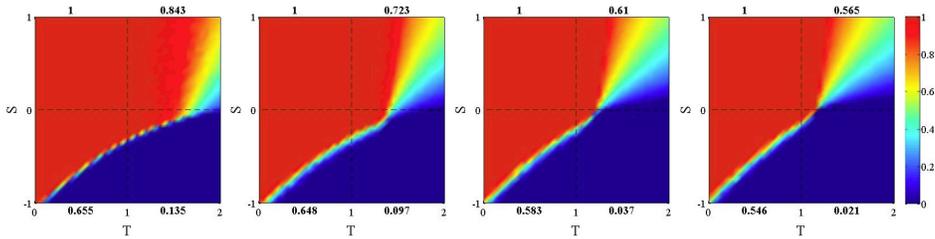} } \protect 
\end{tabular}}
\end{center}
\caption{Asymptotic distribution of strategies in the TS plane in TSNs  using replicator dynamics with
accumulated payoff and asynchronous dynamics.
 Initial density of cooperators is $0.5$ uniformly distributed at random in all cases. Network size is $N=4000$ and
 average degree $\bar k=8$. From left to right: $\omega=0, 10^{-3}, 10^{-2}, 2 \times 10^{-2}$. Values are averages over $100$ independent runs. \label{link-rw-rd}}
\end{figure}

The period of network rewiring is the number of strategy update steps before a node rewiring takes place, and
the frequency $\omega$ is just the reciprocal of this number. 
Figures~\ref{link-rw-rd} show average cooperation levels of all games as a function of the rewiring frequency $\omega$.
The update rule is replicator dynamics with accumulated payoff and asynchronous population evolution.
From left to right: (static case) $\omega=0$, $\omega=10^{-3}$, $\omega=10^{-2}$, $\omega=2 \times 10^{-2}$. 
Values of $\omega$ larger than the latter would cause too fast a rewiring. In actual social networks network dynamics is 
slow or medium-paced, depending on the type of interaction. The above link dynamics  is only intended to qualitatively
model those complex phenomena.
It is clear that, as in the case of BA networks studied in~\cite{antonioni}
cooperation is progressively lost with increasing frequency of rewiring $\omega$. This seems to be a general phenomenon
in all networks: the loss of cooperation is caused by increasingly fast destruction of an individual's environment in the
network. This noise prevents cooperators from forming stable clusters.

\section{Conclusions}
\label{concl}

In this work we have presented a systematic numerical study of standard two-person evolutionary games on two classes of social
network models.  The motivation behind this choice is to make a further step towards more realism in the interacting agents population
structure. The networks have been built according to Toivonen et al. model (TSN)~\cite{toivonen-2006}, one of several social network models used
in the literature and, in part, according to the model proposed by Amaral et al.~\cite{am-scala-etc-2000}. 

Previous investigations have shown that broad-scale network models such as Barab\'asi--Albert (BA) networks are rather favorable
to the emergence of cooperation, with most strategy update rules and using accumulated 
payoff~\cite{santos-pach-05,santos-pach-06,anxo1}. 
Here we have shown that the same is true in general for TSNs and the Amaral model, almost to the same extent as in BA networks. In addition, synchronous
and asynchronous population update dynamics have been compared and the positive results remain true and even better for
the asynchronous case when using imitation of the best update.
We have also presented results for payoff schemes other than accumulated. In particular, we have studied average payoff and various
proportions between the two extreme cases. The general observation is that pure average payoff gives the worst results in terms
of cooperation, as already noted in~\cite{santos-average,Tomassini2007a,attila-average}. When going from average to accumulate payoff
cooperation tends to increase.

Finally, a couple of sources of noise on the evolutionary process have been investigated in order to get an idea about the
robustness of cooperation on TSNs. To introduce strategy errors we have used the Fermi update rule. Cooperation on TSNs
is relatively robust against this kind of noise, in a manner comparable to scale-free graphs~\cite{anxo1}. Of course, when the error
rate becomes high, the behavior resembles to random and cooperation tends to decrease for all nontrivial games. 

With a view to the fact that actual
social networks are never really static, we have designed one among many possible mechanisms to simulate link fluctuations.
When this kind of network noise is present cooperation tends to decrease and to disappear altogether when the network
dynamics is fast enough. Similar effects have been observed in scale-free networks~\cite{antonioni}.

In conclusion, TSNs and also Amaral's networks appear to be as favorable as scale-free graphs for the emergence of cooperation in evolutionary games.
But, with respect to the latter, the additional advantages are that TSNs and Amaral networks are much closer to actual social networks in terms of topological structure and
statistical features. Cooperation can thus emerge and be stable in this kind of networks and probably also on related
models. This is hopefully good news for cooperation among agents in social networks provided that the relationships are
sufficiently stable. However, too much strategy noise or network instability may cause cooperation to fade away as
in any other network structure.

\paragraph{\bf Acknowledgments.} A. Antonioni and M. Tomassini gratefully acknowledge the Swiss National Science Foundation for financial 
support under contract number 200021-132802/1.

\bibliographystyle{ws-acs}
\bibliography{jeux}

\end{document}